\documentclass[a4paper]{article}
\usepackage{amsmath,graphicx}
\usepackage{float}
\usepackage{adjustbox}
\usepackage{makecell}
\usepackage{titlesec}
\titlespacing*{\subsection}{0pt}{1.1\baselineskip}{\baselineskip}
\usepackage{INTERSPEECH2022}

\newcommand\blfootnote[1]{%
  \begingroup
  \renewcommand\thefootnote{}\footnote{#1}%
  \addtocounter{footnote}{-1}%
  \endgroup
}

\title{ACOUSTIC TO ARTICULATORY SPEECH INVERSION USING MULTI-RESOLUTION SPECTRO-TEMPORAL REPRESENTATIONS OF SPEECH SIGNALS}
\name{Rahil Parikh$^1$$^\ast$, Nadee Seneviratne$^1$$^\ast$, Ganesh Sivaraman$^2$, Shihab Shamma$^1$, Carol Espy-Wilson$^1$}
\address{
  $^1$Institute for Systems Research,\\ University of Maryland College Park, USA\\
  $^2$Pindrop Inc., USA}
\email{rahil@umd.edu, nadee@umd.edu}

\begin{document}

\maketitle
\begin{abstract}
Multi-resolution spectro-temporal features of a speech signal represent how the brain perceives sounds by tuning cortical cells to different spectral and temporal modulations. These features produce a higher dimensional representation of the speech signals. The purpose of this paper is to evaluate how well the auditory cortex representation of speech signals contribute to estimate articulatory features of those corresponding signals. Since obtaining articulatory features from acoustic features of speech signals has been a challenging topic of interest for different speech communities, we investigate the possibility of using this multi-resolution representation of speech signals as acoustic features. We used U. of Wisconsin X-ray Microbeam (XRMB) database of clean speech signals to train a feed-forward deep neural network (DNN) to estimate articulatory trajectories of six tract variables. The optimal set of multi-resolution spectro-temporal features to train the model were chosen using appropriate scale and rate vector parameters to obtain the best performing model. Experiments achieved a correlation of 0.675 with ground-truth tract variables. We compared the performance of this speech inversion system with prior experiments conducted using Mel Frequency Cepstral Coefficients (MFCCs).
\end{abstract}

\section{Introduction}
\label{sec:intro}
\blfootnote{$^\ast$ These authors contributed equally to this work.}

Acoustic to articulatory speech inversion is the process of mapping the speech acoustic signal to articulatory parameters associated with the movement of the vocal tract articulators for speech production. When accurately estimated, articulatory movements of the vocal tract can be applied to speech accent conversion \cite{Aryal2014}, speech therapy \cite{cavin2015use}, language learning, Automatic Speech Recognition (ASR) \cite{Kirchhoff2002}, and detection of depression from speech.

Real articulatory data is obtained from subjects using techniques like Electromagnetic Articulometry (EMA), X-ray microbeam, and real-time Magnetic Resonance Imaging (rt-MRI). However these techniques are expensive and time consuming. Obtaining real articulatory data is not practically feasible for real world applications like ASR. Hence, it is essential to develop speech inversion systems that are speaker independent and can accurately estimate articulatory features from the speech signal for any unseen test speaker.

The mapping from acoustics to articulations is known to be highly non-linear and non-unique \cite{Qin2007}. Adding speaker variability to the already challenging problem makes it even more difficult. Based on a comprehensive study of the speech inversion techniques, the speaker dependent techniques can be classified into three categories- 
\begin{itemize}
    \item Codebook based approaches \cite{Atal1978,Ouni2005} in which a codebook of acoustic and corresponding articulatory patterns is constructed from the training data
    \item Analytical approaches involving articulatory models such as Maeda's \cite{Laprie1998}
    \item Statistical modeling (parametric and non-parametric) of acoustic to articulatory mapping like Gaussian Mixture Model (GMM), Mixture density networks (MDN) \cite{Richmond2007}, Hidden Markov models (HMM) \cite{Hiroya2004b}, generalized smoothness criteria \cite{Ghosh2010}, and neural networks \cite{Kirchhoff1999}
\end{itemize}
There have been a few attempts to perform speaker independent speech inversion using deep neural networks (DNN) \cite{Sivaraman2019}. Previously we worked on improving the performance of this baseline DNN based speech inversion system in \cite{Sivaraman2019}. The first improvement was to make the system more robust to noise \cite{Seneviratne2018}. Secondly, we trained a multi-task learning based DNN to train the SI system on multiple articulatory databases in order to increase the generalizability of the TV estimation task \cite{Seneviratne2019}.

Most of the approaches for speech inversion have parameterized the speech signal as mel frequency cepstral coefficients (MFCCs), mel filterbanks, perceptual linear prediction (PLP) coefficients which are commonly used acoustic features for ASR. The homomorphic filtering based cepstral coefficients make sense for speech inversion because they keep only the slowly changing content of the speech signal associated with articulation and discard the fast changing content associated with voice quality and identity. However, with the advent of deep and convolutional neural networks, richer and complete representation of the speech signal has been favored over features like cepstral coefficients. 

The auditory cortex represents speech using multiple spectral and temporal resolutions using cells tuned to different spectral and temporal modulations operating through different frequency specific neural processing channels \cite{Santoro2014}. Thus, the brain performs a higher dimensional representation of the speech spectrum. In the neuroscience literature there are two contending theories of speech perception \cite{Diehl2004}, namely - (1) General Auditory theory \cite{fant1970acoustic, stevens1978invariant} and (2) Motor theory of speech perception \cite{liberman1967perception, liberman1985motor}. The auditory theory argues that speech perception happens solely due to representations of speech in the auditory cortical regions. The motor theory says that speech production information is essential in recognizing speech and contributes to the robustness of the perception. This study brings together the two theories by attempting to estimate articulatory movements using cortical-like multi-resolution spectro-temporal representations of speech. These multi-resolution spectro-temporal representations will be referred to as cortical features in this paper. The cortical features provide a fine grained control over the modulation rates to focus on in the representations. They also provide us valuable knowledge about the relative contributions of different spectral and temporal modulation scales and rates towards estimating the articulatory movements. This knowledge can potentially lead to more robust acoustic feature representations targeted towards optimally capturing the speech articulations while discarding non-speech auditory objects. To the best of our knowledge cortical features have not been used to perform acoustic-to-articulatory speech inversion.

In this paper, we focus on developing a speaker independent speech inversion system using multi-resolution spectro-temporal representations of natural articulatory speech data. We used the multi-speaker XRMB clean speech dataset to perform our experiments. Further details about this dataset is given in section \ref{sec:dataset}. We used the Matlab toolbox developed by the U. of Maryland Neural Systems Laboratory to extract the cortical features of the speech signals. The extracted features' dimensionality was reduced using Higher Order Singular Value Decomposition (HOSVD) before passing to a simple feed-forward neural network to be trained. The development of the model is described in section \ref{sec:sisystem}. The following section \ref{sec:results} will explain the experiments conducted to obtain the optimum parameters and the best performing network along with the results obtained by comparing the model developed with similar other systems which were using MFCCs as acoustic features. Section \ref{sec:conclusion} presents the conclusion.

\section{Dataset Description}
\label{sec:dataset}
To train the model to estimate Tract Variables (TVs), it is necessary to have a dataset which has ground-truth TVs corresponding to its speech utterances. Thus, we used the Wisconsin X-Ray Microbeam database \cite{Westbury1994a} to perform our experiments. The choice of this database also helps to compare the performance of this model with similar experiments conducted earlier using different types of acoustic features.

The XRMB recordings comprise of naturally spoken utterances from 32 male and 25 female subjects along with X-ray Microbeam cinematography of the mid-sagittal plane of the vocal tract with pellets placed at points along the vocal tract. The trajectory data are recorded for the individual articulators: Upper Lip, Lower Lip, Tongue Tip, Tongue Blade, Tongue Dorsum, Tongue Root, Lower Front Tooth (Mandible Incisor), Lower Back Tooth (Mandible Molar). We call these trajectories as pellet trajectories. The miss tracked segments were removed from the database before using it for our analysis.

The TVs specify the salient features of the vocal tract area function more directly than the pellet trajectories \cite{McGowan1994} and are relatively speaker independent. They also provide us a theoretical framework to analyze speech production with the theoretical framework of articulatory phonology. Hence, the pellet trajectories were converted to TV trajectories using geometric transformations as outlined in \cite{Mitra2012}. The transformed XRMB database consists of 21 males and 25 females, with a total of 4 hours of speech data with corresponding 6 TV trajectories. The TVs obtained from the seven pellet trajectories were – Lip Aperture (LA), Lip Protrusion (LP), Tongue Body Constriction Location (TBCL), Tongue Body Constriction Degree (TBCD), Tongue Tip Constriction Location (TTCL) and, Tongue Tip Constriction Degree (TTCD).

\begin{figure}[t]
  \centering
  \includegraphics[width=\linewidth]{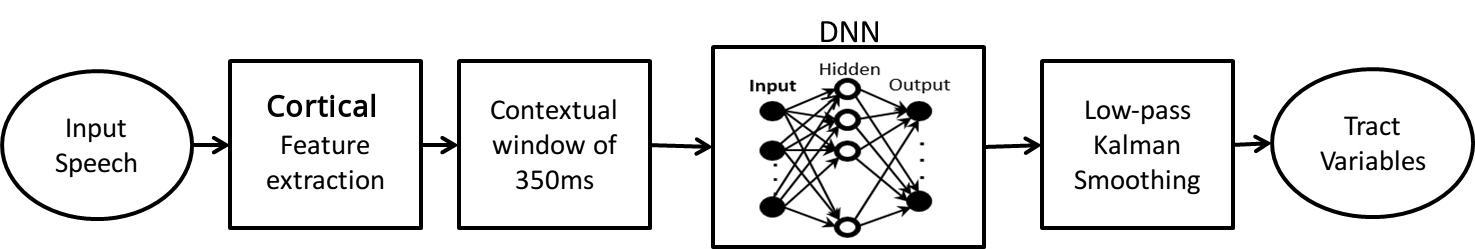}
  \caption{Block diagram of the speech inversion system}
  \label{fig:speech_inversion_system}
\end{figure}

\section{Speech Inversion System}
\label{sec:sisystem}
The steps involved in building the model depicted in Figure~ \ref{fig:speech_inversion_system} to estimate TVs are explained in this section. The details of extracting multi-resolution spectro-temporal features of the speech signal, performing feature reduction on the same, and training and testing the DNN are elaborated below.

\subsection{Feature Extraction}
\label{ssec:subhead}
The cortical features or the multi-resolution spectro-temporal representations \cite{Chi2005} are obtained using an existing toolbox \cite{ru2001multiscale} developed to perform these operations. The audio files are first converted to an auditory spectrogram to emulate the processing performed by the ear \cite{Chi2005}. A bank of constant-Q bandpass filters are used to model the human cochlea. The NSL toolbox uses a cochlear filter bank of 128 filters with center frequencies uniformly distributed along the logarithmic axis, spread over 5.33 octaves. The hair cells and the lateral inhibitory network are modelled as explained in \cite{Wang1995} to yield the auditory spectrogram. The cortical representations are obtained by performing a wavelet transform of this spectrogram with Spectro-Temporal Response Fields (STRFs), which are a bank of 2-Dimensional filters that are tuned to temporal modulations (or rates) which vary from slow to fast, and spectral modulations (or scales) that vary from narrow to broad \cite{Chi1999}. This mimics the role of the primary auditory cortex in computing the temporal and spectral modulation content of the auditory spectrogram \cite{Chi2005}. These are a three dimensional tensor for each frame of the input audio signal. 

In our analysis, we generate the auditory spectrogram using a frame length of 10ms, to correspond to the sampling time of the tract variables signals. The cortical features are generated using STRFs with temporal modulation filters at 2,4,8,16 and 32 Hz for upward and downward moving patters in the spectrogram,and spectral modulation filters at 1,2,4 and 8 cycles/octave. The features are thus a tensor with dimensions $4\times{10}\times{128}$ representing the 4 scale values, 10 rate values and 128 frequency channels, for each time frame of the input audio-sample.
\subsection{Dimensionality Reduction}
\label{ssec:subhead}

The cortical features span over 5120 dimensions for each time frame of 10ms, obtained by 4 scale filters, 10 rate filters, and 128 frequency channels. In such a high dimensional representation, the features are sparse and any machine learning algorithm will suffer from the curse of dimensionality. Fitting a model using such a high dimensional feature would require a huge number of parameters, making it computationally intensive. Hence an efficient dimension reduction method is required to reduce the dimensionality of the features. Unfortunately, conventional dimension reduction methods such as Principal Component Analysis (PCA) fail to discriminate between the contributions of the various feature sub-spaces, such as scales, rates and frequency, and treat all modes similarly. Dimensionality reduction is thus obtained by using principal components from HOSVD \cite{Mesgarani2006}. Since we wish to retain all the information with respect to time, dimension reduction is only performed on the three dimensional scale-rate-frequency tensor. The HOSVD can be obtained by performing SVD on matrices obtained by unfolding the mode of the tensor along which the dimension needs to be reduced \cite{DeLathauwer2000}. A tensor can be decomposed into another tensor of the same dimensions and the left-singular matrices obtained by unfolding each mode of the tensor, as shown in the equation below \cite{Mesgarani2006}:

\begin{equation} \label{tensor_decomposition}
    T=A\times_1{U^s}\times_2{U^r}\times_3{U^f}
\end{equation}

where $U^s$, $U^r$, $U^f$, are the unitary left-singular matrices corresponding to the scales, rates and frequencies, obtained by unfolding each mode of the original tensor, and $A$ is an ordered and orthogonal tensor. The tensor T in equation (\ref{tensor_decomposition}) is 3-Dimensional tensor, obtained by concatenating the absolute values of cortical features of all samples in the training set.  
\begin{figure}[h]
  \centering
  \includegraphics[width=\columnwidth, height=0.7\columnwidth]{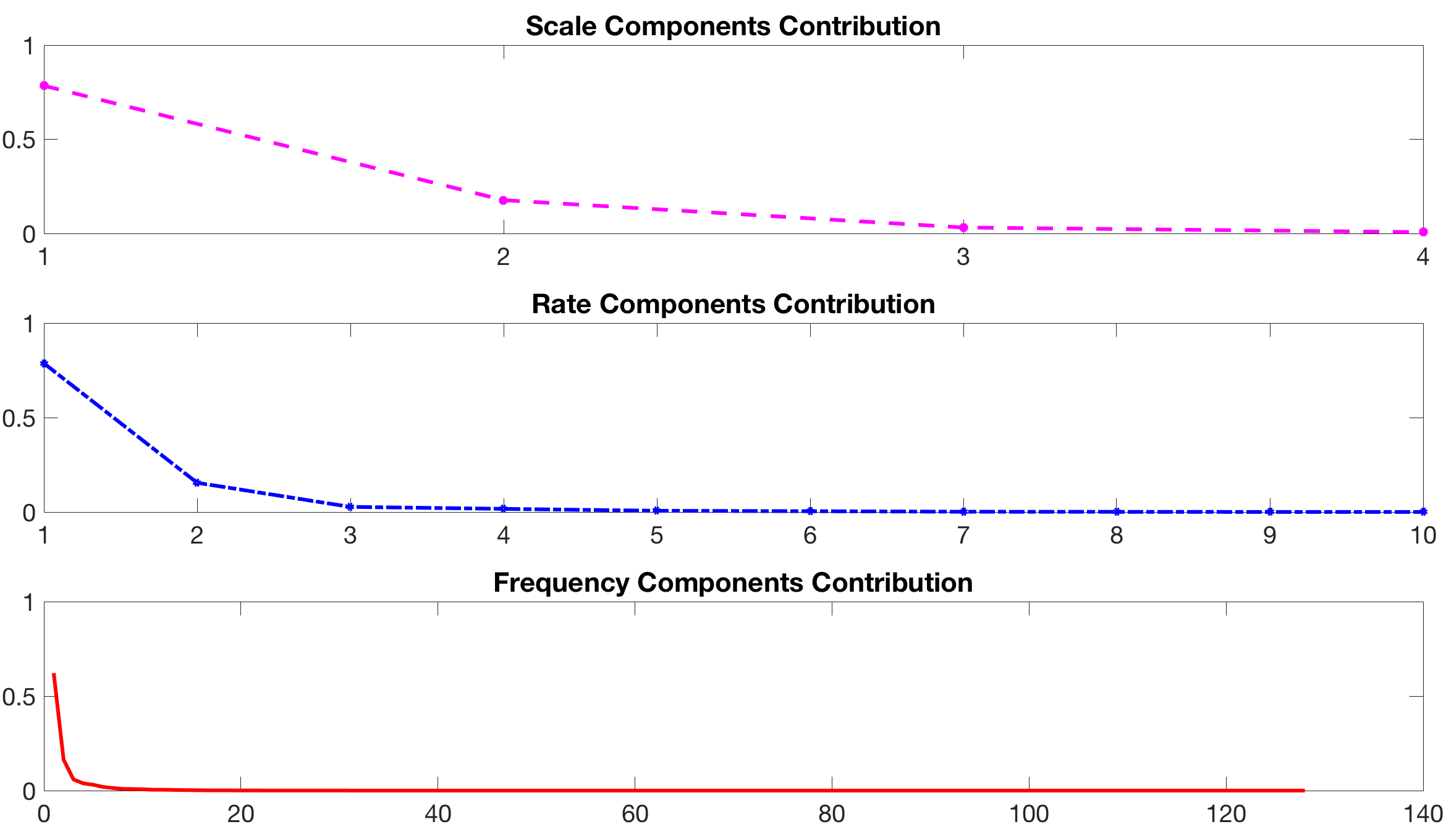}
  \caption{Principal Components Contributions for Scales, Rates and Frequencies}
  \label{fig: PC_Components}
  \end{figure}

To determine the contribution of the principal components (PCs) for each mode, we compute the relative energy of each eigenvalue for the given mode, using the following formula:
\begin{equation}
    \label{pc_cont}
    \alpha_{i,j} = \frac{\lambda_{i,j}}{{\sum_{k=1}^{N_i}\lambda_{i,k}} }
\end{equation}
where $\lambda_{i,j}$ is the corresponding eigenvalue of the $j^{th}$ PC of the subspace $S_i$ and $N_i$ is the dimension of $S_i$. These plots are shown in Figure~\ref{fig: PC_Components}.

Components that contribute significantly are retained, and the tensor is multiplied to the corresponding eigenvectors for the given mode.

For instance, as shown in Figure~\ref{fig: PC_Components} only the first 8 principal components contribute significantly to the information in the subspace. The matrix with eigenvectors corresponding to these 8 eigenvalues are multiplied with the tensor A of dimensions $4\times{10}\times{128}$, to obtain a tensor $A_f$ with dimensions $4\times{10}\times{8}$.
In our experiments, the input tensor of size $4\times{10}\times{128}$ is reduced to a tensor of size $4\times{5}\times{7}$ and size $4\times{6}\times{8}$ for each time frame of an audio-sample. This tensor is then vectorized for each time frame. 

\subsection{DNN Training}
The input layer accepts contextualized cortical features and the output layer (which estimates the TVs) has a dimensionality of 6 nodes. The input dataset was divided into training, development, and testing sets so that the training set has utterances from 36 speakers and the development and testing sets have 5 speakers each (3 males,2 females). Around 80\% of the total number of utterances were present in the training set. The development and testing sets have a nearly equal number of utterances. This allocation of utterances was done in a completely random manner.

In this work we aim to demonstrate that cortical features can be mapped to tract variables even with minimal complexity of the neural network. We thus trained three feed-forward DNNs with 400, 512, and 800 neurons each in all the hidden layers. Similar to prior experiments \cite{Sivaraman2017a}, we observed that as the number of layers increased, the performance increased. After 6 hidden layers the performance on the development set dropped. Hence we chose to fix the number of layers to 6. Thus, DNNs with 6 hidden layers were trained and the model performing better on the development set was chosen as the optimal model.

The DNN was trained on mini batches as the total amount of input features were quite high to be trained as a whole batch. The training was performed to minimize the mean squared error between the actual TVs and the estimated TVs. The Adam optimizer was used for optimizing the network parameters. Since the articulatory trajectories are low-pass in nature, the estimated TVs from the DNN were passed through a Kalman filter to remove additional noise \cite{Sivaraman2017a}

\subsection{Performance Measurement}
The best performing DNN is the one that has the least mean squared error on the validation set. Since the dataset contained articulatory data from multiple speakers, correlation is a more appropriate performance measure than mean squared error. The Pearson Product Moment Correlation (PPMC) between the estimated TV and the corresponding ground-truth TV was computed as the performance evaluation metric.




\section{Experiments and Results}
\label{sec:results}
The cortical feature space was initially reduced to a sub-space with 4,5 and 7 number of scale, rate and frequency principal components respectively to obtain the best performing DNN model out of the models having 400, 512 and 800 nodes in each hidden layer for 6 hidden layers. To investigate the impact of the number of principal components being used in the vectorized features, we performed DNN training on 2 different sets of scale, rate, and frequency components of the input features: $(4,5,7)$ and $(4,6,8)$. Depending on this the size of the input layer was varied as 980 and 1344 nodes respectively. Dropout layers were used to reduce over-fitting. 

\begin{table}[htb]
\caption{Average correlations of estimated TVs for different DNN architectures, using $(4,5,7)$, and $(4,6,8)$ scale, rate and frequency coefficients for the cortical features}
\label{table1}
\begin{adjustbox}{width=\columnwidth}
\begin{tabular}{|c|l|l|}
\hline
\textbf{\# Hidden layer nodes} & \textbf{Cortical\_980 (4,5 7)} & \textbf{Cortical\_1344 (4,6,8)} \\ \hline
400                            & 0.635                          & 0.685                           \\ \hline
512                            & \textbf{0.645}                 & 0.684                           \\ \hline
800                            & 0.632                          & \textbf{0.694}                  \\ \hline
\end{tabular}
\end{adjustbox}
\end{table}




The 6-layer DNN with 512 nodes in each hidden layer yielded the best average correlation of 0.645 on the development set for 4,5 and 7 scale, rate and frequency coefficients.

According to Figure \ref{fig: PC_Components}, the contribution from scale, rate and frequency coefficients of $(4,5,7)$ respectively seems to represent the subspaces well. To get a more accurate representation of the input features, the rate and frequency components were increased to 6 and 8. The same experiments were repeated using 0.1 dropout for the input layer and 0.2 dropout for the subsequent layers. The results obtained on the development set are summarized in Table \ref{table1}.


It can be seen that preserving more principal components from scale and frequency axes improves the correlation by $7.6\%$.
The Table \ref{table3} summarizes and compares the average correlation values for each of the 6 TVs for different input features for speech inversion - Cortical\_980, Cortical\_1344, and MFCCs.

\begin{table}[!htb]
\caption{Average Correlations for Estimated TVs (defined in Section \ref{sec:dataset})}
\label{table3}
\begin{adjustbox}{width=\columnwidth}
\begin{tabular}{|c|c|c|c|c|c|c|c|c|}
\hline
\textbf{Feature} & \makecell{\textbf{Context}\\ \textbf{frames}} & \textbf{LA}  & \textbf{LP}  & \textbf{TBCL} & \textbf{TBCD} & \textbf{TTCL} & \textbf{TTCD} & \makecell{\textbf{Average}\\\textbf{Correlation}} \\ \hline
Cortical\_980 (4,5,7) & 7  & 0.723        & 0.494        & 0.751         & 0.534         & 0.523         & 0.768         & \textbf{0.632}               \\ \hline
Cortical\_980 (4,6,8) & 7 & 0.758        & 0.516        & 0.789         & 0.589         & 0.576         & 0.822         & \textbf{0.675}               \\ \hline
MFCC & 17 & 0.856 & 0.613 & 0.866 & 0.745 & 0.707 & 0.907 & \textbf{0.782}
                 \\ \hline
\end{tabular}
\end{adjustbox}
\end{table}

\section{Conclusion}
\label{sec:conclusion}
In this work we bring together the General Auditory Theory and the Motor Theory of speech perception. We develop a speech inversion system using the multi-resolution spectro-temporal `cortical features' which are critical in the perception of sound in mammals. Experiments performed in the paper prove that these representations can estimate the articulatory features (TVs) with a correlation of 0.675 with ground-truth TVs. 

The HOSVD dimensionality reduction preserves the slow varying components of speech which are ideal for estimating articulatory features.  Since MFCCs directly correspond to capturing vocal tract variations and phonetic characteristics of speech signals, the correlations are much higher compared to the results obtained from our experiments in this paper. Moreover, the experiments with MFCC features had used a feature concatenation of 17 frames as against 7 frames for the cortical features. However the proper choice of rate and scale vectors has a significant impact to the correlation. The rates of 4, 8, 16 and 32 Hz capture the syllabic rate of the speech signals which corresponds to the tract variable trajectories. Hence with the appropriate parameter and feature dimensionality selection it can be seen that it is feasible to use the rich representation of the speech signals provided by cortical features to capture speech articulations. 

In the future we plan to explore a joint optimization of the dimensionality reduction and the articulatory feature estimation so that the dimensionalities are optimally reduced for the speech inversion problem.

\bibliographystyle{IEEEtran}

\bibliography{template}


\end{document}